\def\p{\partial}
\def\a{\alpha}

\def\D{\Delta}
\def\e{\varepsilon}

\def\T{\theta}

\def\s{\sigma}
\def\t{\tau}

\def\Ra{\Rightarrow}

\documentclass[12pt]{article}

\usepackage{a4wide}
\usepackage{graphicx}

\usepackage{amsfonts}

\begin{document}
\title{Nonlinear analysis of a simple model of temperature evolution 
in a satellite}
\author{Jos\'e~Gaite, Angel~Sanz-Andr\'es and Isabel~P\'erez-Grande\\[2mm]   
{\small IDR/UPM, ETSI Aeron\'auticos, Universidad Polit\'ecnica de Madrid,}\\
{\small Pza.\ Cardenal Cisneros 3, E-28040 Madrid, Spain} }
%

\maketitle

\begin{abstract}
We analyse a simple model of the heat transfer to and from a small satellite
orbiting round a solar system planet. Our approach considers the satellite
isothermal, with external heat input from the environment and from internal
energy dissipation, and output to the environment as black-body radiation.
The resulting nonlinear ordinary differential equation for the satellite's
temperature is analysed by qualitative, perturbation and numerical methods,
which show that the temperature approaches a periodic pattern (attracting
limit cycle).  This approach can occur in two ways, according to the values of
the parameters: (i) a slow decay towards the limit cycle over a time longer
than the period, or (ii) a fast decay towards the limit cycle over a time
shorter than the period. In the first case, an exactly soluble average
equation is valid.  We discuss the consequences of our model for the thermal
stability of satellites.
\end{abstract}

\section{Introduction}
\label{intro}


The design process of any spacecraft necessarily includes a thermal analysis,
performed to guarantee that both the bus and the payload are kept along the
whole mission within the appropriate temperature range
\cite{therm-control}. The thermal
environment of a satellite in orbit is very harsh, dominated by vacuum
conditions which make difficult to find a way of cooling down the components
of the spacecraft. Every single element of the satellite has to be thermally
analysed to verify that its requirements are fulfilled. To do it, it is
necessary a rather detailed design of the spacecraft, as it implies to
calculate thermal loads and couplings between elements. This is usually
carried out by commercial software tools based on numerical methods.


During the generic trade-off study, early in the design process, that is, when
the concept of the spacecraft is still open, it is very convenient to assess
the mean temperature of the satellite and its variation by means of analytical
tools. This method provides valuable information about the feasibility of the
spacecraft from the thermal point of view as well as the possibility of easily
carrying out parametric analyses to study the influence of the different
variables involved in the problem.

The purpose of the present study is to analyse the thermal behaviour of a
small compact satellite spinning in a low orbit round a solar system planet
(the Earth, say) as a function of time, using analytical or semi-analytical
tools. In orbit, the spacecraft is subjected to periodic heat loads and it is
expected to reach a thermally periodic state. Apart from obtaining the
temperature oscillations once reached this periodic state, this analytical
approach allows us to study the convergence of the satellite temperature to
the periodic behaviour starting from different initial conditions, as it can
occur from launch conditions or orbital manoeuvres.

The results are obtained by applying the energy balance equation to the
satellite, taking into account both the external heat loads, i.e. the solar
irradiation, the planet albedo (solar irradiation reflected on the planet) and
the infrared (IR) radiation from the planet, as well as the internal
dissipation.

We begin with the formulation of the energy balance equation as a
non-dimensional ordinary differential equation (ODE), which is nonlinear and
non-autonomous. Then, we consider an autonomous ODE that plays the role of an
average of the actual equation. That autonomous equation is exactly soluble
and guides our intuition of the behaviour of the actual equation and helps in
its analysis. This analysis is carried out with qualitative, perturbation and
numerical methods. Finally, we present our conclusions, regarding the design
of satellites.

\section{Nonlinear ODE for the satellite temperature}
\label{sec:1}

Let us assume that the satellite is approximately isothermal, so we
characterize it by its temperature $T$. The heat input consists of an external
flow coming from the space environment, in particular, the solar irradiation,
the planetary albedo and the planetary IR radiation, and the internal heat due
to the equipment dissipation. The solar irradiation and the planetary albedo
have periodic variation, according to the position of the satellite on its
orbit.  On the contrary, the planetary IR radiation and the internal heat
operate at constant rate.

Let us assume that the solar irradiation changes from constant to vanishing
(when the satellite is in the planet's shade), whereas the albedo depends on
the angle between the sunlight and the vertical to the satellite (and is much
smaller in absolute value).  Furthermore, we assume that the albedo vanishes
for one fraction of a period whereas the satellite's night (when it is in the
planet's shade) lasts for another (smaller) fraction.  We denote the heat rate
due to solar irradiation by $\dot{Q}_s$, and the maximum albedo heat rate by
$\dot{Q}_a$ (at the satellite's noon).  In addition, let the constant rate
source have power $\dot{Q}_c$ (the sum of the equipment dissipation and
planetary IR radiation).  Then, the energy balance equation that yields the
temperature $T$ is
\begin{eqnarray*}
C \, \dot{T}(t) = 
\dot{Q}_s\,f_s(\nu t) + \dot{Q}_a\,f_a(\nu t) + 
\dot{Q}_c 
- A\epsilon\s\,{T(t)}^4;\\
f_s(x) = 1, \;0 \leq x \leq x_1\;{\rm or}\; 1-x_1 \leq x \leq 1;\;
f_s(x) = 0, \; x_1 < x < 1-x_1;\\
f_a(x) = \cos (2\pi x), \;0 \leq x \leq x_2 \;\textrm{or}\; 
1-x_2 \leq x \leq 1;\; f_a(x) = 0, \;x_2 \leq x \leq 1-x_2;\\
f_{s,a}(x) = f_{s,a}(x - 1), \;x \geq 1. 
\end{eqnarray*}
Here $C$ is the satellite's thermal capacity, $\nu$ the orbital frequency, $A$
is the satellite's surface area, $\epsilon$ its emissivity, and $\s$ the
Stefan-Boltzmann constant.  The values of $x_1$ and $x_2$ are smaller than one
half and they determine the fractions of the period with vanishing albedo or
sunshine, respectively. In principle, $x_2 = 1/4$ and $x_1 > x_2$ (we assume
that the albedo vanishes for half a period whereas the satellite's night is
shorter: we take it as one fifth of the period, that is, $x_1 = 2/5$, in an
example below).

We can write this equation in non-dimensional form by defining
$a=A\epsilon\s/(C\nu)$, $k_s=a^{1/3}\,\dot{Q}_s/(C\nu)$,
$k_a=a^{1/3}\,\dot{Q}_a/(C\nu)$, $k_c=a^{1/3}\,\dot{Q}_c/(C\nu)$, and
non-dimensional temperature variable $\T=a^{1/3}\,T$ and time variable $\nu t$
(which we still denote $t$ for notational simplicity). Then,
\begin{equation}
\dot{\T}(t) = k_c + k_s\, f_s(t) + k_a\,f_a(t) - {\T(t)}^4.
\label{ODE}
\end{equation}
Unfortunately, this nonlinear ODE cannot be reduced to a quadrature; but we
can deduce its relevant properties, nevertheless.

We note that, if we remove the oscillating terms, the resulting equation
\begin{equation}
\dot{\T}(t) = k - {\T(t)}^4
\label{ODE2}
\end{equation}
(where $k = k_c$) is immediately reduced to a quadrature, which can be
integrated analytically. Moreover, we can deduce the qualitative behaviour of
the solutions in a straightforward way: there is one fixed point, $\T_{\rm eq}
= {k}^{1/4}$, and it is stable.  It is the equilibrium temperature, at which
the heat input and the radiation output balance one another. When the
temperature is close to it, $\T = {k}^{1/4} + \D \T$, we obtain the linear ODE
$$
\dot{\D \T}(t) = - 4 {k}^{3/4} {\D \T(t)},
$$
with solution
\begin{equation}
{\D \T(t)} = {\D \T(0)}\, \exp(-4 {k}^{3/4} t).
\label{linODE-sol}
\end{equation}
Therefore, ${\D \T}$ halves in a time ${\D t} = (\ln 2/4)\, {k}^{-3/4} = 0.17
/{\T_{\rm eq}}^{3}$; and the larger is ${\T_{\rm eq}}$, the shorter it takes
to reach equilibrium. We have checked that the exact solution of 
Eq.~(\ref{ODE2}) is well approximated by the linear equation solution 
(\ref{linODE-sol}). 

We remark that Eq.~(\ref{ODE2}) can be naturally connected with
Eq.~(\ref{ODE}) if we assume that $k$ is the average of $k_c + k_s\, f_s(t) +
k_a\, f_a(t)$ over one period, namely, $k = k_c + 2 x_1 k_s + k_a/\pi$ (rather
than $k = k_c$).  We explore the consequences of this connection in the next
section.

With $k = k_c + 2 x_1 k_s + k_a/\pi$, reverting to physical variables, 
we have 
\begin{equation}
T_{\rm eq} = \left[\frac{\dot{Q}_c + 2 x_1\, \dot{Q}_s +
\dot{Q}_a/\pi}{A\epsilon \s}\right]^{1/4}.
\label{Teq}
\end{equation}
Note that it does not depend on the parameters $C$ and $\nu$, associated to 
the time derivative of the temperature in the ODE in physical variables.

\section{Solution of the nonlinear ODE for the temperature evolution}
\label{sec:2}

In this section, we first present a qualitative analysis, which allows us to
prove the existence of an attracting limit cycle. To obtain the transient
behaviour and the properties of the limit cycle, we employ perturbation
theory.  Finally, we perform a numerical analysis of the ODE, to check the
results of the preceding methods and to obtain concrete and more precise
numerical results.

\subsection{Qualitative analysis} 

Equation (\ref{ODE}) is non-autonomous and therefore equivalent to an
autonomous system of two ODE's, namely, the system formed by Eq.~(\ref{ODE})
and the trivial equation $\dot{t} = 1$.  The generic behaviour of an
autonomous system of two first-order ODE's is to have ``simple'' attracting
sets, namely, equilibrium points or limit cycles, because of the
Poincar\'e-Bendixson theorem \cite{Hi-Sm,Andro,Drazin}.  Given the nature of
our problem, we can also apply the theory of ODE's with periodic coefficients
(see, e.g., Ref.\ \cite{Arnold}): we can reduce the dynamics in the
$(t,\T)$-plane to the cylinder $[0,1) \times (0,\infty)$. Furthermore, we can
consider $\T$ as the radial coordinate and $2\pi t$ as the angular coordinate
of a plane with the origin excluded (equivalent to the mentioned cylinder).
Given that there cannot be fixed points (since the equation $\dot{t} = 1$
forbids it), the Poincar\'e-Bendixson theorem implies that the only possible
attractors are limit cycles, with the period imposed by the heat input.

The Poincar\'e-Bendixson theorem states that a curve solution of a
two-dimensional autonomous ODE that has no singularities (e.g., fixed points)
and is contained in a compact domain for all $t \geq 0$ approaches a limit
cycle. We use a consequence of this theorem: given two concentric closed
curves limiting an annular region that is free of singularities and such that
the vector of derivatives on the two curves points towards the inside of the
region, there exists a limit cycle in this annulus.

To prove the uniqueness of the limit cycle, we appeal to Dulac's criterion for
an annular region: if the autonomous system
$$\dot{x} = P(x,y),\;\dot{y} = Q(x,y)$$ is such that the divergence $\p_x P +
\p_y Q$ has constant sign in an annular region, then this region contains at
most one limit cycle.  The proof of this criterion follows from Green's
theorem \cite{Andro}.

It is easy to check that Eq.~(\ref{ODE}) is such that $\T < \T_{\rm min} =
{k_c}^{1/4} \Ra \dot{\T} > 0$ and $\T > \T_{\rm max} = {(k_s + k_a +
k_c)}^{1/4} \Ra \dot{\T} < 0$, for all $t$.  Note that $\T_{\rm min}$ and
$\T_{\rm max}$ correspond to the equilibrium temperatures with {\em constant}
minimum or maximum heat input, respectively.  Therefore, the trajectories in
the plane with $\T$ as the radial coordinate and $2\pi t$ as the angular
coordinate which begin inside the annulus defined by those two temperatures
are confined in it, and there is (at least) one limit cycle (with an
oscillation in $\T$ confined to take place within those values).  Furthermore,
this limit cycle is unique, because $\textrm{div} (k_c + k_s\, f_s(t) +
k_a\,f_a(t) - {\T}^4,1) = -4\,\T^3 < 0.$

\begin{figure}
\centering{\includegraphics[width=8.8cm]{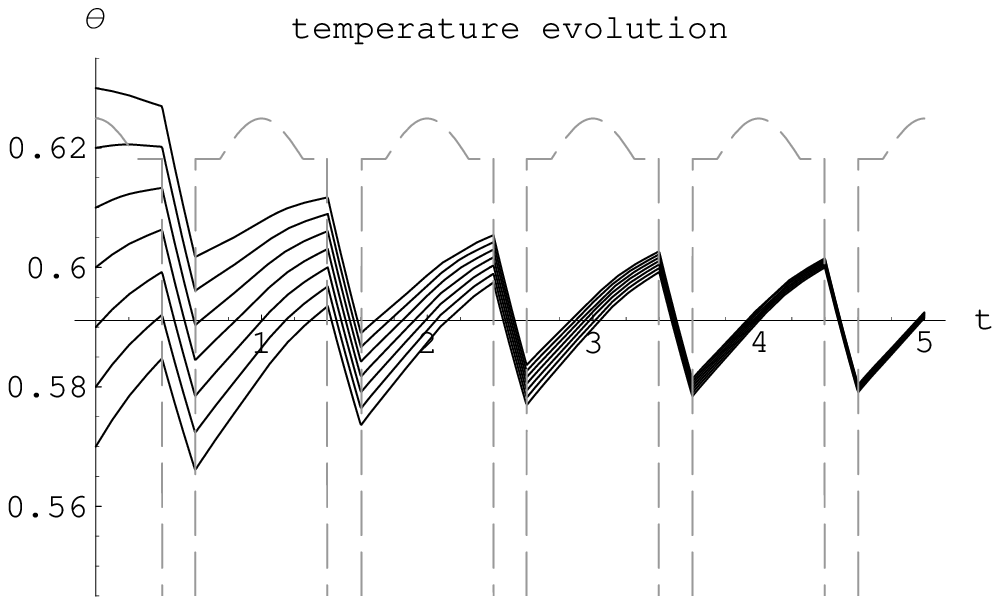}\hbox{\phantom{aa}}
\includegraphics[width=6.8cm]{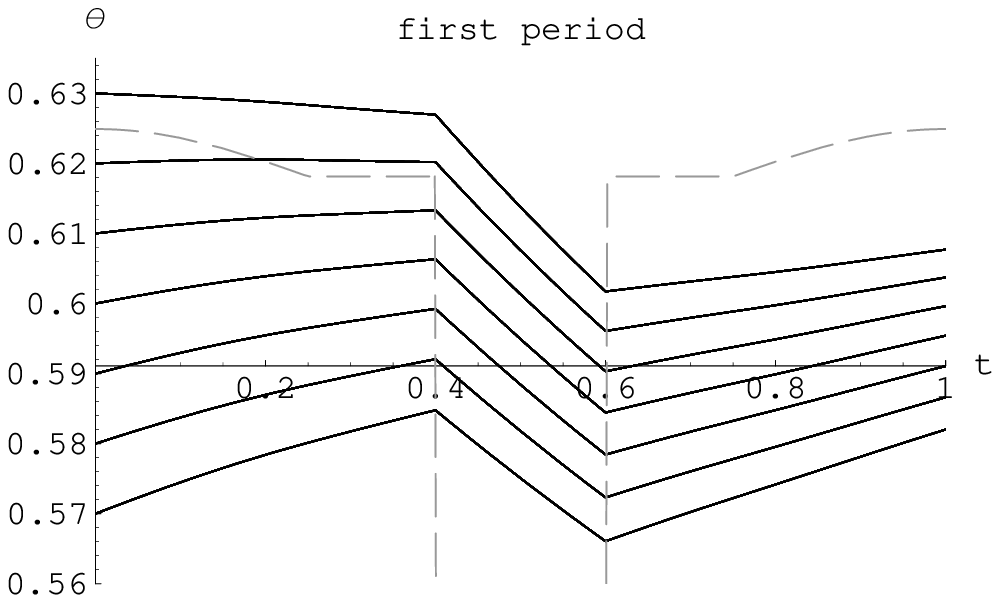}}
\caption{Results of the numerical analysis of the ODE with 
$k_s = 0.13$, $k_a = 0.007$ and $k_c = 0.016$, explained in the text: (left)
convergence to the attractor; (right) detail of the first period.  Note how
the trajectories change from increasing to decreasing, and viceversa, as they
cross the dashed line ${\T_{\rm lim}(t)}$ (the bottom of this line, at
${k_c}^{1/4} = 0.36$, is below the plot regions).}
\label{plot}
\end{figure}

The trajectories inside the annulus are alternately increasing and decreasing,
as the right-hand side of Eq.~(\ref{ODE}) changes sign. To be precise, if we
define the periodic function
$$
{\T_{\rm lim}(t)} =
\left[ k_c + k_s\,f_s(t) + k_a\,f_a(t)
\right]^{1/4},
$$
with minimum and maximum values $\T_{\rm min}$ and $\T_{\rm max}$, 
respectively, the sign of $\dot{\T}(t)$ changes when the trajectories 
cross it. This is represented in Fig.~\ref{plot}. 

\begin{figure}
\centering{\includegraphics[width=7.5cm]{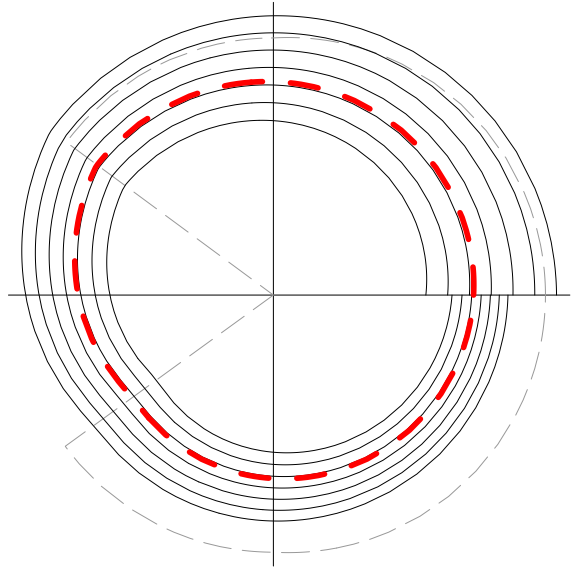}}
\caption{Polar plot of the trajectories in Fig.~\ref{plot} (right): 
the radius is $\T-0.5$ (subtracted to display well-separated lines)
and the angle is $2\pi t$. The limit cycle is the bold dashed line.}
\label{polar-plot}
\end{figure}

Using $\T$ as the radial coordinate and $2\pi t$ as the angular coordinate,
Fig.~\ref{plot} becomes the polar plot in Fig.~\ref{polar-plot}. This polar
plot is more useful to represent convergence to the limit cycle in a standard
way.

We are interested in two questions: (i) finding the features of the limit
cycle and the rate of convergence to it; (ii) analysing how this limit cycle
and the rate of convergence depend on the constants $k_c, k_s, k_a$. These
questions cannot be answered by a qualitative analysis, so we turn to other
methods.

However, let us note that some more qualitative information can be
obtained from a comparison with the non-oscillating Eq.~(\ref{ODE2}): if
$\T_{\rm eq} = {k}^{1/4} = (k_c + 4 k_s/5 + k_a/\pi)^{1/4}$ is sensibly smaller
than one, the convergence time ${\D t} = {\T_{\rm eq}}^{-3}$ is large, so that
it is consistent to consider Eq.~(\ref{ODE2}) as an averaged equation. To be
precise, Eq.~(\ref{ODE2}) can be derived as an equation for the mean
temperature in Eq.~(\ref{ODE}) by averaging it over one period.  Therefore,
the evolution is given by a long-time decay to $\T_{\rm eq}$ and a short-time
oscillation about that long-time behaviour (as in Fig.~\ref{plot}).

Thus, it seems convenient to consider Eq.~(\ref{ODE}) as a perturbation of
Eq.~(\ref{ODE2}), namely, to consider the time-dependent (oscillating)
functions $f_{a,s}$ as a perturbation. Then, we can employ standard
perturbation methods \cite{Ben-Or}. We do so in the next section.

\subsection{Perturbation theory}

As we have explained above, our intention is to take advantage of the simple
solution of Eq.~(\ref{ODE2}) by taking it as the basis of a perturbation
scheme. Therefore, we write Eq.~(\ref{ODE}) as 
\begin{equation}
\dot{\T}(t) = k + \e \, f(t) - {\T(t)}^4,
\label{ODE-pert}
\end{equation}
where $k = k_c + 2 x_1\, k_s + k_a/\pi$, $f(t) = k_c - k + k_s\, f_s(t) +
k_a\,f_a(t)$, and $\e$ is a formal perturbation parameter to be set to one at
the end. Note that $f(t)$ is defined such that it has vanishing average over a
period and, therefore, it represents the deviations about the mean value $k$.
Furthermore, the solution of Eq.~(\ref{ODE-pert}) must fulfill the initial
condition $\T(0) = \T_{\rm in}$. We assume an expansion of $\T(t)$ of the form
$$
\T(t) = \sum_{n=0}^{\infty} \e^n \T_n(t),
$$
and substitute it into Eq.~(\ref{ODE-pert}). Equating to zero the successive
powers of $\e$, we obtain the basis equation and 
an infinite set of linear ODE's, namely,
\begin{eqnarray}
\dot{\T}_0(t) = k - {\T_0(t)}^4,\\
\dot{\T}_1(t) = f(t) - 4 {\T_0(t)}^3 \T_1(t), \label{T1}\\
\dot{\T}_2(t) = - 6 {\T_0(t)}^2 {\T_1(t)}^2 - 4 {\T_0(t)}^3 \T_2(t),
\label{T2}\\
\dots\quad\quad\dots\quad\quad\dots\phantom{aa}  \nonumber
\label{linODE-pert}
\end{eqnarray}
The initial condition for $\T_0(t)$ is $\T_0(0)=\T_{\rm in}$, while the
remaining equations fulfill $\T_n(0) = 0$.

These ODE's can be solved subsequently, namely, we can solve first the
equation for $\T_0(t)$ and, then, we can solve the equation for $\T_1(t)$ and
so onwards. Of course, the equation for $\T_0(t)$ is Eq.~(\ref{ODE2}), which
we have already solved.  The following equations are just first-order linear
inhomogeneous equations, which are soluble by quadratures
\cite{Ben-Or}. However, their solution involves complicated integrals that
cannot be made analytically. In particular, Eq.~(\ref{T1}) has the solution
$$
{\T}_1(t) = \frac{1}{I(t)} \int_0^t f(\t) I(\t)\, d\t,\quad
I(t) = \exp\left[4\int_0^t {\T_0(\t)}^3 d\t \right]; 
$$
but $I(t)$ has no analytic expression. 

To overcome this difficulty and obtain analytic expressions, we can use
instead of the exact function $\T_0(t)$ the approximated function given by
Eq.~(\ref{linODE-sol}), namely, $\T_0(t) = \T_{\rm eq} + (\T_{\rm in} -
\T_{\rm eq}) \exp(-4 {k}^{3/4} t)$. Thus, the integral in $I(t)$  becomes 
a sum of integrals of exponential functions, which yield exponential functions
again.  In the large-$t$ limit (when $t \gg {\T_{\rm eq}}^3$), the
exponentials decay and the value of $I(t)$ further simplifies to $I(t) =
\exp\left( 4 {k}^{3/4} t \right)$. Then,
\begin{equation}
{\T}_1(t) = \int_0^t f(\t) \exp\left[- 4 {k}^{3/4} (t-\t) \right] d\t =
\int_0^t f(t-\t) \exp\left[- 4 {k}^{3/4} \t \right] d\t.
\end{equation}
Furthermore, in the large-$t$ limit and given that $f(t)$ is periodic, we can
extend the upper integration limit in the latter integral from $t$ to
$\infty$. Thus, we have a periodic function, namely, an approximation to the
limit cycle (past the transient regime). We can obtain a general expression of
this periodic function through Fourier analysis.

\subsubsection{Fourier analysis}

Let us expand $f(t)$ in Fourier modes as
$$
f(t) = \sum_{m = -\infty}^{\infty}f_m\, e^{2\pi i m t}.
$$
Since $f(t)$ is real and we assume that it is symmetric with respect to 
$t= 0$, the coefficients $f_m$ are real and $f_{-m} = f_m$; 
in addition, $f_0 = \int_{0}^{1} f(t)\,dt = 0$. 
The other Fourier coefficients are given by 
\begin{equation}
f_m = \int_{0}^{1} f(t)\, e^{-2\pi i m t}\,dt\,.
\label{Fourier-coef}
\end{equation}
It is straightforward to solve for ${\T}_1(t)$:
\begin{equation}
{\T}_1(t) = \sum_{m = -\infty}^{\infty} 
\frac{f_m\, e^{2\pi i m t}}{2\pi i m + 4 {k}^{3/4} }
= 2 \sum_{m = 1}^{\infty} f_m\,
\frac{4 {k}^{3/4} \cos(2\pi m t) + 2\pi m\, \sin(2\pi m t)}%
{4 \pi^2 m^2 + 16 {k}^{3/2}}
\, .
\label{Fourier-sol}
\end{equation}

If we substitute the form of $f(t) = k_c - k + k_s\, f_s(t) + k_a\,f_a(t)$
into Eq.~(\ref{Fourier-coef}), we obtain
$$
f_m = k_s \frac{\sin(2\pi m x_1)}{\pi m} + 
\frac{k_a}{2} 
\left[\frac{\sin(2\pi (m+1) x_2)}{\pi (m+1)} + 
\frac{\sin(2\pi (m-1) x_2)}{\pi (m-1)} \right],
$$
when $m \neq 0$.
With $x_2 = 1/4,$ 
$$
f_m = k_s \frac{\sin(2\pi m x_1)}{\pi m} - \frac{k_a}{\pi(m^2 -1)} 
\cos\left(m \frac{\pi}{2}\right) .
$$

On the other hand, independently of the form of $f_m$, if 
the convergence time ${\D t} = {\T_{\rm eq}}^{-3}$ is large, we have that 
$4 {k}^{3/4} \ll 2\pi$, so we can neglect in 
Eq.~(\ref{Fourier-sol}) the terms with $k$ and write the crude approximation
\begin{equation}
{\T}_1(t) = 
2 \sum_{m = 1}^{\infty} f_m\, \frac{\sin(2\pi m t)}{2 \pi m}
\, .
\label{Fourier-sol-approx}
\end{equation}
This approximation is equivalent to neglecting in the right-hand side of
Eq.~(\ref{T1}) the second term with respect to the first one, that is to say,
it implies that ${\T}_1(t)$ follows $f(t)$ with no delay. We can use it
to provide an estimation of the amplitude of the oscillations, with the
following simple method. This estimation can be useful in the conceptual
design of the satellite. 

\begin{figure}
\centering{\includegraphics[width=9.5cm]{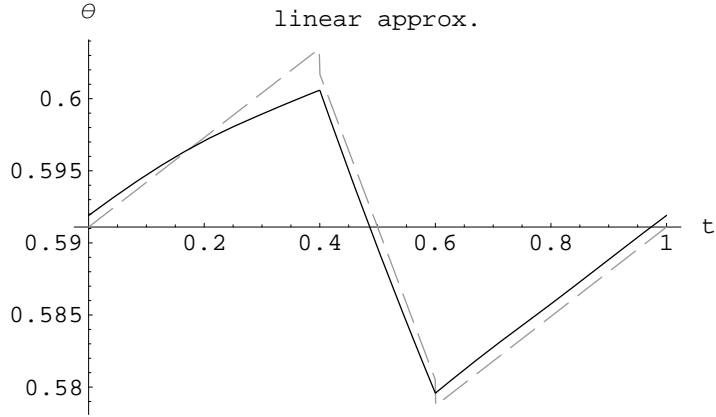}}
\caption{Linear approximation to the first-order perturbative 
limit cycle $\T_{\rm eq} + \T_1(t)$, 
such that it is linear by pieces (dashed line), 
compared to the real limit cycle (full line).}
\label{lin_approx}
\end{figure}

The largest variations of the slope of ${\T}(t)$ and of its approximation
${\T}_1(t)$ take place at the discontinuities of $f(t)$, namely, at
$t=x_1,\,1-x_1$ (in the first period). $f(t)$ changes sign at those times and
the periodic function ${\T}_1(t)$ has its maximum and minimum there. In
between, we can take $f(t)$ constant and, therefore, ${\T}_1(t)$ linear.
Hence, we compute the slopes at two convenient points, namely, $t=0,\,1/2$,
where ${\T}_1(t)=0$, according to Eq.~(\ref{Fourier-sol-approx}). We use this
information to calculate the maximum and minimum of ${\T}_1(t)$.  The slope at
$t=0$ is $f(0) = (1-2x_1) k_s + (1-1/\pi) k_a$, so the maximum is
\begin{equation}
{{\T}_1}_{\rm max} = x_1\,f(0) = 
x_1\left[(1-2x_1) k_s + (1-1/\pi) k_a \right].
\label{max-A}
\end{equation}
Then, with $k_s = 0.13$, $k_a = 0.007$ and $k_c = 0.016$ (values to 
be justified in Sect.~\ref{num-sol}),
${{\T}_1}_{\rm max}(x_1=0.4) = 0.4 \times 0.031 = 0.012$.  
The slope at $t=1/2$ is $f(1/2)
= -2x_1 k_s - k_a/\pi$, so the maximum is 
\begin{equation}
{{\T}_1}_{\rm max} = -(1/2 -x_1)\,f(1/2) = 
(1/2 -x_1)\left[2x_1 k_s + k_a/\pi \right]. 
\label{max-B}
\end{equation}
In particular,
${{\T}_1}_{\rm max}(x_1=0.4) =
(0.5-0.4) \times 0.106 = 0.0106$.  Both values agree sufficiently. Since
${\T}_1(t)$ in Eq.~(\ref{Fourier-sol-approx}) is an odd function, the minimum
is at $t=1-x_1 = 0.6$ and its value is the negative of the maximum.
A graphical comparison of the present approximation with the actual limit
cycle is displayed in Fig.~\ref{lin_approx}.

Note that Eqs.~(\ref{max-A}) and (\ref{max-B}) coincide if $k_a = 0$.  In
fact, the preceding method relies on the particular form of $f(t)$, namely, it
is close to a step function. This is the reason why the method yields a good
result. Of course, while still using the approximation leading to
Eq.~(\ref{Fourier-sol-approx}), the preceding method is improved by the
formula
$$
{\T}_1(t) = \int_0^t f(\t)\,d\t\,,
$$
which demands a little more work to yield the maximum value of $\T_1$.  Better
approximations to $\T_1$ are possible by considering $k$ in
Eq.~(\ref{Fourier-sol}), that is to say, by considering the displacement of
${\T}_1(t)$ [${\T}_1(0) \neq 0$]. In a different sense, the approximation is
also improved by carrying on the perturbation scheme to the second order.

\subsubsection{Second order approximation}

Eq.~(\ref{T2}) is similar to Eq.~(\ref{T1}), if we consider $\T_0 = {\T_{\rm
eq}}$ and (the square of) the first order solution as the forcing
term. Therefore, a similar reasoning leads us to the existence of a periodic
solution for $\T_2(t)$ in the large-$t$ limit, which can be obtained by
Fourier analysis. The result is
\begin{equation}
{\T}_2(t) = -6 \,{k}^{1/2} \sum_{p = -\infty}^{\infty} 
\frac{e^{2\pi i p t}}{2\pi i p + 4 {k}^{3/4} }
\sum_{m = -\infty}^{\infty} 
\frac{f_m\,f_{p-m}}{2\pi i p - 4 \pi^2 m(p-m) + 16 {k}^{3/2} }
\, .
\label{Fourier-sol2}
\end{equation}

We can see that the perturbation series is a power series in the Fourier
coefficients of the forcing term and its convergence depends on the magnitude
of this term. Given that this term is proportional to $k_s$ and $k_a$, the
convergence of the series is improved when these constants are small. This
condition is related to the one that we found for Eq.~(\ref{ODE2}) to hold as
an average equation, namely, that $k = k_c + 2 x_1\, k_s + k_a/\pi$ is
small. Thus, it is consistent with Eq.~(\ref{ODE2}) as the basis of the
perturbation scheme.

\subsection{Numerical solutions}
\label{num-sol}

We use a standard numerical method to integrate Eq.~(\ref{ODE}), for given
values of $k_s, k_a, k_c$. Let us see what values are adequate for a small
satellite that is orbiting round the Earth and which can be considered
isothermal.

We can take the satellite to be a cube of $0.5$~m side and mass of 50~kg.  Let
us assume that it is mostly covered by solar cells. Then, suitable values for
the absorptivity and emissivity are $\a = 0.8$ and $\epsilon = 0.7$,
respectively. We assume the satellite's average specific thermal capacity to
be 0.2~W~h~kg$^{-1}$~K$^{-1}$ (corresponding to a composition of alluminium
with some plastic).  An adequate value for the orbital frequency is
(1.5~h)$^{-1}$.  Therefore, $C \nu = 50\,\textrm{kg}\; 0.2\,
\textrm{W\,h\,kg}^{-1}\,\textrm{K}^{-1} (1.5 \,\textrm{h})^{-1} = 
6.67$~W/K and
$a=A\epsilon\s/(C\nu) = 1.5\,\textrm{m}^{2}\,0.7\;
5.7\, 10^{-8}\,\textrm{W}\,\textrm{m}^{-2}\,\textrm{K}^{-4}/
(6.67\, \textrm{W/K})  
= 8.99\, 10^{-9}\,\textrm{K}^{-3}$.

The solar irradiation heat input is the product of the solar constant, the
absorptivity and the projected area (a quarter of the real area); namely,
$\dot{Q}_s = 1370\, \textrm{W}\,\textrm{m}^{-2}\,
0.8\,[6(0.5\,\textrm{m})^2/4] = 411$~W. Therefore, $k_s = a^{1/3}\,
\dot{Q}_s/(C \nu) = 0.128$.  The albedo is very variable, and to calculate
$\dot{Q}_a$, it is necessary, in addition, to consider the {\em visibility
factor}. We take an average value of $\dot{Q}_a = 0.05\,\dot{Q}_s$ so $k_a =
0.0065$.

The constant heat input is the sum of the equipment dissipation plus the
planetary IR irradiation input.  The equipment dissipation power is due to the
transformation of the incoming irradiation power. If we assume that the solar
cells cover 80\% of the satellite's surface and their efficiency is about
10\%, and we take into account that the solar irradiation only holds for a
fraction $2x_1 = 4/5$ of the period, we deduce that the power dissipated is
about 30~W.  The planetary IR irradiation input can be estimated by applying
the energy balance to the planet (Earth). This balance yields an IR radiation 
flux of about 7\% of the solar constant. We can take the absorptivity in the
IR to be similar to the emissivity $\epsilon = 0.7$. Hence, the 
planetary IR irradiation is about 23~W and $\dot{Q}_c = 53$~W.

Summarizing, we can take $k_s = 0.13$, $k_a = 0.0070$ and $k_c =
0.016$. Furthermore, we consider seven values of the initial condition $T(0)$
($t=0$ is the noon) uniformly distributed in the interval 273~K $\leq T \leq$
300~K, corresponding to nondimensional $0.57\leq \T \leq 0.63$.  These values
are included between $\T_{\rm min} = {k_c}^{1/4} = 0.36$ and $\T_{\rm max} =
{(k_s + k_a + k_c)}^{1/4} = 0.625$. The results are plotted in
Fig.~\ref{plot}, for the time interval $0 \leq t \leq 5$, which is sufficient
to show convergence to the limit cycle.

Indeed, in our example, we are in a situation in which $\T_{\rm eq} = (k_c +
4k_s/5 + k_a/\pi)^{1/4} = 0.591$ is sufficiently small to apply the averaged
equation (\ref{ODE2}). The convergence time is ${\D t} = {\T_{\rm eq}}^{-3} =
4.8$. Fig.~\ref{plot} clearly shows that the evolution is given by an
oscillation with unit period superimposed to a slower and approximately
exponential convergence in a time ${\D t}$.

To measure precisely the attractor characteristics, we choose the initial
condition $\T(0) = \T_{\rm eq} = 0.591$ and we let the numerical integration
up to $t = 10$.  The resulting solution has $\T(10) = 0.592$, which is
slightly different from $ \T_{\rm eq}$ (although the difference is
inappreciable with two-digit precision). Of course, the solution with $\T(0) =
0.592$ is the limit cycle. It has local minimum and maximum amplitudes 0.58
and 0.60 at $t = 2/5$ and $t = 3/5$, respectively. Those extreme values are
quite close, considering the interval defined by $\T_{\rm min}=0.36$ and
$\T_{\rm max}=0.625$ (see Fig.~\ref{plot}). The limit cycle is plotted in
Fig.~\ref{lin_approx}.

\section{Discussion}
\label{sec:3}

Regarding the problem of heat transfer in a satellite, the main conclusions
that we can draw from our analysis are the following. First of all, as regards
the temperature stability, it holds in the sense of convergence to the limit
cycle behaviour, guaranteed by the application of the Poincar\'e-Bendixson
theorem and Dulac's criterion for an annular region.  Moreover, the
convergence is exponential.  

The limit cycle behaviour has a time dependence related to the periodic heat
input, namely, approximately similar to its integral, but somewhat displaced.
To be precise, the limit cycle consists, in one period, of a
temperature-growing phase when the solar heat input is on (as is the albedo),
and a relaxation phase in which the temperature falls to adjust to the
constant input $\dot{Q}_c$ (although this phase ends long before the
temperature approaches the corresponding equilibrium value).  The total
amplitude of the temperature oscillation is relatively small in the example
that we have studied.  This example, with realistic values of the
non-dimensional constants $k_s, k_a, k_c$ for a low-orbit small satellite
(reasonably small values, in particular), shows that the attractor is
approached exponentially but in a time reasonably larger than the period. This
time is simply given in terms of $k_s, k_a, k_c$ by ${\D t} = k^{-3/4}$, where
$k = k_c + 2x_1k_s + k_a/\pi$.  Moreover, the mean temperature in the limit
cycle is ${\T_{\rm eq}} = {k}^{1/4}$.

It is useful to compare the mean temperature $\T_{\rm eq}$ with the
oscillation about it. The former depends on $k$, namely, the average of $k_c +
k_s f_s(t) + k_a f_a(t)$, whereas the latter depends on $f(t)$, which is the
oscillation of $k_s f_s(t) + k_a f_a(t)$ ($k_c$ is a constant).  Thus, the
temperature oscillation is independent of $k_c$ (the constant heat input).
This constant is bound to be quite smaller than $k_s$, for physical reasons,
since the equipment dissipation power is due to the transformation of the
incoming solar irradiation power, and the IR irradiation input is a fraction
of the solar input.  On the other hand, $k_a \ll k_s$, and we assume that $x_1
> 1/4$.  In conclusion, the main contributions to $k$ and, therefore, to
$\T_{\rm eq}$ are due to $k_s$. If the temperature oscillation is estimated by
Eqs.~(\ref{max-A}) or (\ref{max-B}), the main contribution to it also seems to
be due to $k_s$; except for the factor $1-2x_1$, which can be small.  Let us
first note that Eq.~(\ref{max-A}) yields a larger value than Eq.~(\ref{max-B})
if $x_1 > 1/(2\pi)$, which always holds. Focusing on Eq.~(\ref{max-A}), we
note that the coefficient $1-2x_1$ of $k_s$ is smaller than the coefficient of
$k_a$. When $x_1$ approaches $1/2$, we only have the $k_a$-term: then the
satellite is always under the solar irradiation, which does not oscillate; in
consequence, the temperature oscillation is only due to the albedo and,
therefore, is depressed.

Let us express the temperature oscillation given by Eq.~(\ref{max-A}) in
physical variables:
$$
\left[T - T_{\rm eq}\right]_{\rm max} = a^{-1/3}\,{{\T}_1}_{\rm max} =
x_1\frac{(1-2x_1) \dot{Q}_s + (1-1/\pi) \dot{Q}_a}{C\nu}\,.
$$
Contrary to the expression of $T_{\rm eq}$ in Eq.~(\ref{Teq}), this expression
depends on $C$ and $\nu$ but it does not depend on $A,\,\epsilon$ and $\s$. We
expect both expressions to be very useful for conceptual thermal design.

Of course, in the numerical example in Sect.~\ref{num-sol} we have not
considered every orbital circumstance and we have just intended to find a set
of sensible values for the constants. More information on satellite design and
the space thermal environment can be found in the literature
\cite{therm-control}.
It is worthwhile to discuss briefly here possible changes of the values of
$k_s$ and $k_c$. While the ratio $k_c/k_s \simeq 0.1$ is adequate for a small
satellite, the absolute values of $k_s, k_c$ can be amply changed; for
example, by increasing the period (which we have taken as 1.5~h in our
case). It is easy to see that both $k_s$ and $k_c$ are proportional to the
$4/3$rd-power of the period. Thus, if we increase the period by a factor of
ten, say, the constants increase by a factor of $10^{4/3}$, so that ${\D t} =
{(k_c + 2x_1k_s + k_a/\pi)}^{-3/4}$ decreases by a factor of ten, becoming
smaller than one (the period). In this situation, the convergence to the limit
cycle is very fast. In fact, it is appropriate to consider the opposite
approximation to the temperature evolution, namely, during the decay to the
limit cycle, the temperature only changes slightly along this limit cycle.

Further analysis could be based on a more refined model than the isothermal
model that we have used. In particular, a {\em two-node} model has been
studied by Guerra, P\'erez-Grande and Sanz \cite{IDR}. The two ODE's for the
two-node model are equivalent to an autonomous system of three ODE's.  As is
well known, such system can have chaotic behaviour \cite{Drazin}.  However, we
expect that the situations that are well approximated by the one-node model
that we have analysed do not exhibit chaotic features and, therefore, can be
conveniently studied, and provide, in addition, a basis for this very simple
model.

\subsection*{Acknowledgments}

We thank Antonio Barrero-Gil for conversations.


\end{document}